# A Model-Driven Methodology for Automotive Cybersecurity Test Case Generation

Stefan Marksteiner
ITS Global Research & Technology
AVL List GmbH
Graz, Austria
Email: stefan.marksteiner@avl.com

Peter Priller
ITS Global Research & Technology
AVL List GmbH
Graz, Austria
Email: peter.priller@avl.com

*Abstract*—Through international regulations (most prominently the latest UNECE regulation) and standards, the already widely perceived higher need for cybersecurity in automotive systems has been recognized and will mandate higher efforts for cybersecurity engineering. The UNECE also demands the effectiveness of these engineering to be verified and validated through testing. This requires both a significantly higher rate and more comprehensiveness of cybersecurity testing that is not effectively to cope with using current, predominantly manual, automotive cybersecurity testing techniques. To allow for comprehensive and efficient testing at all stages of the automotive life cycle, including supply chain parts not at hand, and to facilitate efficient third party testing, as well as to test under real-world conditions, also methodologies for testing the cybersecurity of vehicular systems as a black box are necessary. This paper therefore presents a model and attack tree-based approach to (semi-)automate automotive cybersecurity testing, as well as considerations for automatically black box-deriving models for the use in attack modeling.

## I. Introduction

Modern vehicular systems have a high degree of complexity and possess a plethora of embedded, networked cyber-physical systems that include their own processing units. They are therefore considered systems-of-systems. Because of this complexity (dozens of control systems with a combined number of several hundreds of thousands lines of program code) and manifold communication interface (e.g. wireless - cellular, Bluetooth, WiFi, V2X, RFID, cellular, etc.) vehicles have a big surface for cyber attacks. Because of the physical attributes of vehicles (mass, speed, direct interaction with potentially many people), their huge numbers (vehicle fleets), and their development towards autonomous systems, such attacks can develop an enormous threat potential. In order to secure these systems and facilitate their resilience against cyber attacks, it is crucial to prevent as many vulnerabilities in development and isolate threatened single components as much as possible. In the concept phase, this can be achieved through architectural measures (e.g. [1]) and secure platforms, while during implementation appropriate development processes, best practices (including e.g. code reviews) have to be adhered – ISO/SAE standard (DIS) 21434 [2] describes a structured approach for security engineering. However, the secure development of automotive systems has to be verified and validated, which, comprehensively for a complete system means testing after system integration to cover as many vulnerabilities as possible, including one that emerge only through the combination of the system parts. A new regulation by the United Nations Economic and Social Council - Economic Commission for Europe (UNECE) explicitly mandates this kind of testing [3]. In order to keep up with the higher demand of automotive cybersecurity testing that comes with this development, the testing processes will have to be industrialized by automation.

Because of distributed development, configuration and maintenance in the automotive supply chain one has none or only very limited access to knowledge about concrete, internal system implementation. Particularly testing for (or by) other actors than the manufacturers (e.g. users, authorities or assemblers of third party equipment) lacks often of data about the system architecture, configuration files, source code etc. of the system-under-test (SUT). This means that testing in this case would mean black box testing. Therefore, this leads to the problem of selecting suitable test cases (i.e. attack vectors) for a cybersecurity test (as black box system test). The high complexity, which leads to a high combinatorial diversity, creates an extremely high number of possible

This research has received funding from the program "ICT of the Future" of the Austrian Research Promotion Agency (FFG) and the Austrian Ministry for Transport, Innovation and Technology under grant agreement No. 867558 (project *TRUSTED*) and within the the ECSEL Joint Undertaking (JU) under grant agreement No. 876038 (project *InSecTT*). The JU receives support from the European Union's Horizon 2020 research and innovation programme and Austria, Sweden, Spain, Italy, France, Portugal, Ireland, Finland, Slovenia, Poland, Netherlands, Turkey. The document reflects the author's view only and the Commission is not responsible for any use that may be made of the information it contains.
NB: appendices, if any, did not benefit from peer review. A preprint of this paper has been deposited on ArXiv.

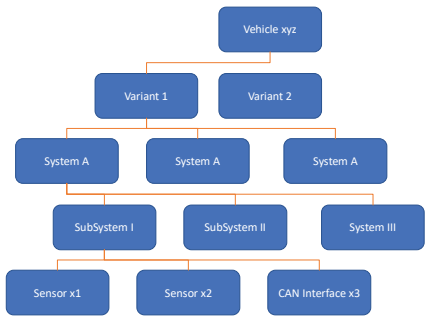

Figure 1. An SUT model

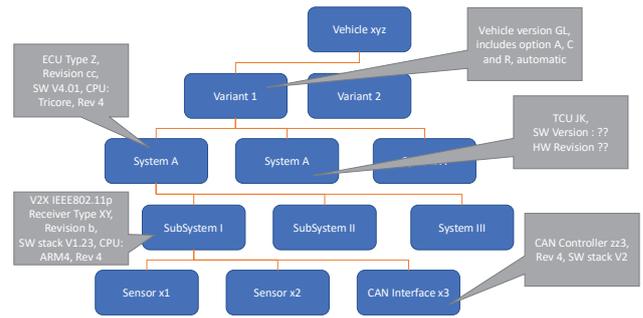

Figure 2. SUT model with component identification guesses by fingerprinting

test cases that cannot be exhaustively tested in a feasible period of time. This paper therefore presents a system for automatically deriving SUT models and reducing their variants in a black box situation and combining it with attack vector generation, automatically generating a feasible number of security test cases for black box-testing of automotive systems. This paper is structured as follows: Section II illuminates an approach for deriving an SUT model or limiting the model variants, respectively. Section III displays an appropriate method to generate attacks on these models and Section V discusses the workflow of a testing system based on the methods described before. Section VI, eventually, concludes the paper.

*A. Related Work and Contribution*

This paper sets on various existing methods for determining a model of an SUT (see Section II), reducing the potential test space (Section II-A), deriving attacks and generate test cases based on attack trees (or similar - see Section III). The related work for each part of the proposed system is cited in the respective sections; high-level overviews of model-based and black-box security testing can be found at [4] and [5], respectively.

The main contribution of the work is the combination of the different methods in an overall workflow (see Section V) and the iterative testing approach (see Section IV).

## II. TEST MODEL DERIVATION

In order to analyse a SUT for suitable attack vectors, one needs a model of the system first (Figure 1 shows a generic example). The more precise the model, the easier to find and/or map specific attack vectors to it. For a black box situation, this means to use any available interface as access vector for system reconnaissance:

- Wireless interfaces (WiFi, Bluetooth, etc.)
- Wired user interfaces (USB, etc.)
- Diagnosis interfaces (OBD-II, etc.)
- Direct wiretapping on buses (CAN, LIN, etc.)

At each of these interfaces, reconnaissance strategies (particularly component fingerprinting and model learning – see next sections) should be performed to receive a model, as concise as possible, of the SUT that can be used for attack vector generation (see Section III); using different angles of reconnaissance combined together in an overall model could yield a more complete description of the SUT.

One approach to automatically identify the components (most prominently electronic control units - ECUs) inside a vehicle is fingerprinting. A tool that can serve as a role model is the network reconnaissance (particularly port scanning) tool *Nmap*. This tool also provides operating system detection by sending probes to an SUT and comparing the answers with a database. Due to subtle differences in the implementation of network standards, the tool can detect a likely operating system on the target SUT using up to 16 TCP, UDP and ICMP probes generated in sequences and differing in e.g. the TCP options [6]. Using similar databases for automotive systems, the principle can transfer to a vehicular system either:

- Passively by listening to vehicular communications
- Actively by sending message sequences to the components and analyzing the answer

Features to detect ECUs in automotive systems are e.g. the clock skew (the difference between an ECUs local clock $C_i$ and an assumed *true* clock $C_{true}$) [7] or a combination of mean, standard and average deviation, skewness, kurtosis, RMS amplitude, as well as minimum and maximum values [8]. The second method is more universal but needs, in contrast to the first, special measurement equipment [9]. On vehicular bus systems access is straight-forward, which effectively reduces that to a software problem on how to create meaningful sampling and test sequences to reliably determine the vast variety of different components inside a vehicular system. In order to industrialize the testing we propose to

1) Use combinatorial logic to generate a sensible set

2) Alter the system under test through controlled conditions (which means e.g. to fingerprint the system, update it using a controlled process yielding a controlled state and fingerprinting it again to highlight the delta behavior).

For other systems inside a vehicle (e.g. an infotainment system running on Android or QNX), more generic methods for IT systems (e.g. Nmap) or interface-specific methods (e.g. for Bluetooth [10]) may be used. Also, artificial intelligence-based methods can be utilized for automotive fingerprinting, e.g. artificial neural networks (ANNs) can yield satisfactory results in CAN channel and ECU detection [11] and deep convolutional neural networks (CNN) have displayed very good performance in terms of detection rate for vehicles [12].

Depending on the respective interface of detection, the component is attributed to a certain network of the car, allowing for creating a topology graph as in Figure 1. Once the device mapping has been performed, a query on a vulnerability database that may include external sources, for example the *Common Vulnerbilites and Exposures(CVE)*-based *National Vulnerability Database (NVD)*[1], matches vulnerabilities to the identified components, whereby the severity of the latter serves as a means for estimating the cost for attacking the respective component as used in an attack vector for a test case (see Section III). Figure 4 displays an example model with matched vulnerabilities. In the context of this paper, this resembles a static method for test generation (using previously known attacks).

A different approach is to use model learning and model checking for dynamic attack vector generation (yielding previously potentially unknown attacks). Instead of deriving a model of components and comparing them with a vulnerability database, a behavioral model is derived from the SUT. An approach for learning a behavior model combining model-based test case generation and a recurrent neural network is presented at [13]. A learned automaton can also be learned for usage with symbolic execution [14]. However, this approach needs a white-box component inside the SUT to yield sensible test cases. In order to overcome this, there is an approach that uses abstract automata learning to learn an abstract model of the SUT, and performs conformance testing on the SUT using fuzzing techniques for concretizing the test input [15]. On the other hand, models can be subjected to a symbolic model checker, which is an already elaborated field [16], that would yield attacks for components that could also serve as a basis for an attack graph. SysML(-Sec) (state) models, for instance, can be translated to ProVerif [17], which is a symbolic model checker for security analysis [18].

[1]https://nvd.nist.gov/vuln

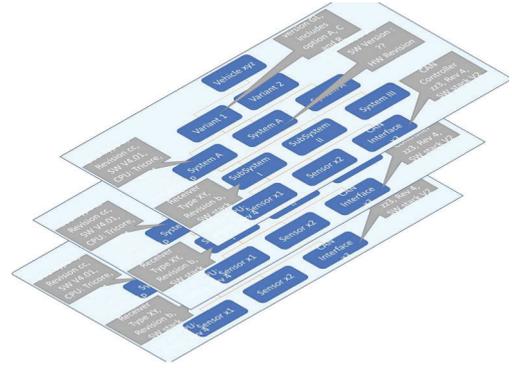

Figure 3. SUT model variants displayed as parallel planes

*A. Coping with Variants*

Functional testing of mechatronic systems has entered automotive engineering some decades ago. Verification and Validation (V&V) methods and processes have since then considerably matured, covering all the way from model-based testing, to x-in-the-loop testing, to component, system, vehicle, proving ground and fleet testing [19]. However, there are important factors which sets cybersecurity testing apart from established V&V methods. First, it never finishes: even if a system has successfully passed all tests before being released, testing needs to continue with each new threat and exploit coming up throughout the whole life cycle of the product. This not only requires active monitoring of such new threats, but also ways to understand impact, select the systems potentially vulnerable to it, derive appropriate test cases and scenarios, and having access to systems under test. Second, many of today's cyber-physical systems (CPS) in cars have significantly different life cycles compared to previous generations, as users begin to expect quick updates similar to e.g. their smart phones. The IoT paradigm of having smart systems (vehicle, house, personal, etc.) communicate with each other even increases this need for fast updates, as a change in one domain might trigger the need to change automotive systems as well to keep compatibility. Third, products become more and more individualized. While mass-production was key in the rise of the automobile during the last century, intelligent production fulfills now more and more the rising demands of users for deep customization. While a wealth of options is preferable during the sales process, it can quickly become a nightmare for updates and maintenance. Combinatorial explosion of variants leads to exponential rise in testing efforts, especially if OEMs cannot limit the number of versions existing in parallel. This is especially hard if manufacturers have not fully implemented OTA (over the air) updating of each software component in the vehicle. Here, the classic way of updating (flashing) in certified car workshops is definitely a bottleneck in achieving version consistency

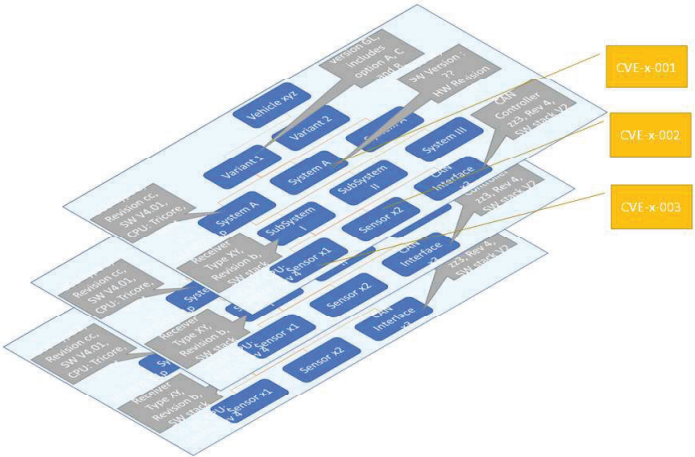

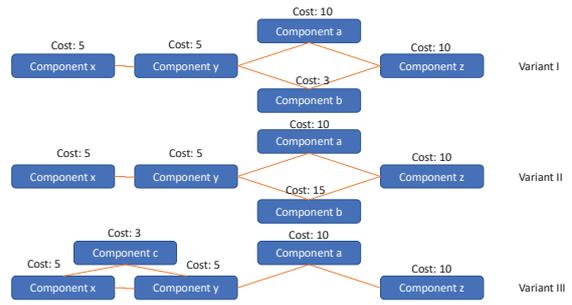

Figure 5. Attack with Costs for Three Model Variants

Figure 4. Matching SUT models with vulnerabilities for attack vector generation

across a fleet.

One way or another, the number of variants will increase over time. Vehicle owners expect OEM's to provide support for their vehicles to extend for significantly longer time than e.g. manufacturers support their smart phones. Variant management (e.g. [20]) ameliorates the problem, but still the traditional value chain in the automotive business (OEM/tier-1/tier-2) and thus many stakeholders are a challenge. An efficient Cybersecurity testing methodology therefore needs to support engineers not only in identifying relevant attacks, test cases and scenarios. It also needs to find the collective of variants of a system under test where this threat applies to.

Another challenge for model-driven cybersecurity testing is posed by uncertainties in system models. In one possible scenario, the test engineer has narrowed down a selection of a few candidate models which may or may not correspond to the SUT (see Figure 3). This situation can be resolved by creating a common attack tree from the superposition of all candidate models. Fingerprinting data is accumulated during the testing process, which is then used to exclude contradictory model variants and their corresponding paths in the common attack tree. As an additional optimization step, we can prioritize attack vectors that contain model elements within the difference set of the candidate models. In this way, these elements will be either fingerprinted or it will be shown that these elements cannot be reached by any known attack vectors, which will render the different model variants indistinguishable. However, this refinement of the testing process is only sensible if it is actually shown to reduce the overall cost of security testing. We may explore the use of AI Planning techniques [21] to determine the optimal or near-optimal next testing step based on estimated attack costs and future gains.

In the above case, we assumed that one of the candidate test models in fact corresponds to the SUT. Generally, test models could be erroneous, incomplete or totally absent. The elicitation of security models can be seen as a Bounded-Synthesis-problem. There are examples of model-variant-based model learning techniques in which a SAT solver is used to continuously exclude alternative variants [22]. Alternatively, Formal Methods like Alloy [23], [24] could be used to keep track of all possible models of a given size that correspond to the accumulated fingerprinting data. This option is highly dependent on our ability to discover additional components within the SUT and is considered future work.

### III. ATTACK VECTOR GENERATION

Once a system model in form of a topology graph is derived, one can start to generate a security test case in the form of an attack vector that would aim for a specific target.

At a sufficient level of detail, the identified system components will be (automatically) correlated with databases of known vulnerabilities[2]. This yields a model containing references to the vulnerabilities (see Figure 4). In a next step, potential previously unknown weaknesses, derived by e.g. static binary analysis [25], [26] or dynamic binary analysis [27] can be examined. The subsequent goal is to set up a security verification test case for a distinct attack target (e.g. the speed control in an adaptive cruise control system). This means to find a path from a defined outside interface (e.g. Bluetooth, V2X, USB or OBD-II) through the model (and all plausible variants thereof) to the targeted ECU and function using utilizing the matching vulnerabilities at the respective intermediate nodes. The vulnerabilities should be rated, which yields a cost of attack (COA) for each individual node. This allows for generating an attack tree [28], [29], or, to boost the possibilities of the graph by being able to model state, other graphical methods like Petri nets [30]–[32]. A series of test cases for that specific target will consist of cases that each represent an attack vector beginning with the one with the lowest cost. The SUT passes the test if none of the tests cases in the series yields a successful attack, where all vectors are chosen that fall short of a defined minimum security value (MSV) that is deemed sufficient. The MSV is in that context defined as a value that represents the threshold for the cheapest allowed path through an attack tree (or other representation, respectively), to be determined by the testing entity.

---
[2]e.g. NVD, Auto-ISAC, etc.

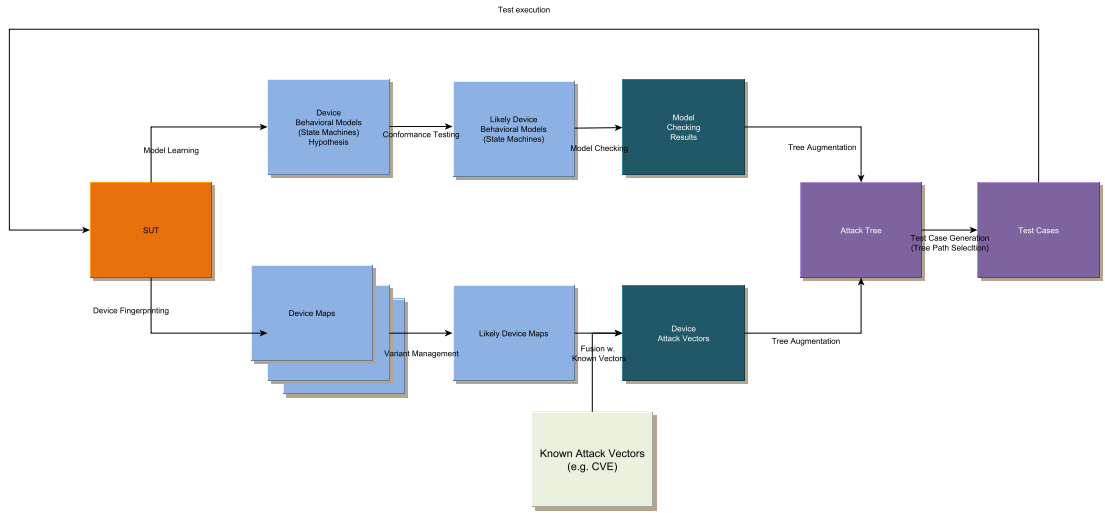

Figure 6. Testing Workflow

Subsequently, a mitigation strategy can be simulated in the model, which alters the path costs and therefore would influence the cost for the cheapest path. When also pricing the mitigation strategies, it is trivial to optimize them to raise the COA above the threshold for the MSV, given that appropriate strategies and pricing values for their implementation are present. When lacking a value for one or more mitigation strategies, it can either be assumed that no strategy is present (independent from whether this is the case or the strategy merely lacks a value for its cost) or a speculative value can be assumed that is inversely proportional to the COA, reasoning that the raising security level of already secure components is more effort-intense than securing a component that yet completely lacks any security measures[3]. Figure 5 displays an attack tree for three variants of a model. The shortest path COA for Variant I is 23, for Variant II is 30 and for Variant III 28. Assuming an MSV of 30 and that boosting low-level security components is cheaper, it would be optimal to suggest a mitigation strategy for node b in Variant I and for node c in Variant III[4]. The testing approach can be set to practical use using a framework as described in e.g. [33].

## IV. ITERATIVE MODEL IMPROVEMENT

The model-derivation, attack vector generation and mitigation is iterative in two aspects:

- Attacks and system access allow for refining the model of the SUT;
- Mitigation strategies influence the model and raise the COA.

The first point means that a) once an attack is successful, it is trivial to gain information about the attacked system (e.g. if system access is given by executing *uname -a* on a Linux system) - but also unsuccessful attacks might provide insight to system components and protective measures - and b) once a system (e.g. an ECU) could be accessed, adjacent systems and networks (e.g. another bus system a gateway ECU is connected to) could be further assessed and the model refined, as well as contradicting variants excluded. The second point refers to the security-raising impact of mitigation strategies by raising the cost of a component where the strategy is imposed. If affecting the lowest-costing path it would raise the overall COA, whereby the attack tree will have to be recalculated including the altered values[5]. To secure the system as desired, the most economical simulated mitigation strategies might be introduced into the model until the system's security is above the MSV. This yields a cost/benefit optimized list of security measures to

---

[3]It is evident that this reasoning is an oversimplification. However, the purpose is to give a rough estimation in a situation where no data is present.

[4]However, to suggest a mitigation strategy that assures a COA above the MSV for all plausible variants alike, only invariant components could be examined. This would mean to select a strategy for components x or y

[5]Simply considering the next more expensive part falls to short as a) it might also be altered and b) the former most expensive part might still be below the threshold

reach a sufficient minimum security level, which can be used as a security requirements catalog.

## V. Discussion

The set of methods outlined assemble a framework for security testing by setting them into a workflow using both device fingerprinting and model learning approaches to automatically derive models of the SUT, reducing the number deviating variants (using model-based variant management or conformance testing, respectively) and using external sources and model checking to identify potential attack vectors. These vectors are used to annotate a graph-based structure (e.g. an attack tree) to derive potential attack paths on the overall system. The different paths of this tree form complete test cases, which are not only used for testing the SUT but also for the COA and MSV determination. See Figure 6 for an overview.

## VI. Conclusion

This paper outlined a set of methodologies to industrialize automotive cybersecurity testing by automation. It displayed approaches for static (through fingerprinting) and dynamic (through model learning) SUT model generation and subsequent attack vector generation using graph-theoretical methods, combined with an iterative approach. The paper therefore showed the feasibility of automating automotive cybersecurity testing by presenting an approach to combining that methodological set to a means for automated cybersecurity test case generation for automotive systems. The efficiency enhancement to be expected will help to tackle the challenge set by the requirements from the UNECE regulation and similar legal obligations.

## References

[1] P. Papadimitratos, L. Buttyan, T. Holczer, E. Schoch, J. Freudiger, M. Raya, Z. Ma, F. Kargl, A. Kung, and J. Hubaux, "Secure vehicular communication systems: design and architecture," *IEEE Communications Magazine*, vol. 46, no. 11, pp. 100–109, 2008.

[2] International Organization for Standardization and Society of Automotive Engineers, "Road Vehicles – Cybersecurity Engineering," International Standard, International Organization for Standardization, ISO/SAE Standard "21434", 2020.

[3] United Nations Economic and Social Council - Economic Commission for Europe, "UN Regulation on uniform provisions concerning the approval of vehicles with regard to cyber security and of their cybersecurity management systems," United Nations Economic and Social Council - Economic Commission for Europe, Tech. Rep., 2020.

[4] M. Felderer, P. Zech, R. Breu, M. Büchler, and A. Pretschner, "Model-based security testing: a taxonomy and systematic classification," *Software Testing, Verification and Reliability*, vol. 26, no. 2, pp. 119–148, 2016.

[5] R. Lachmann, M. Felderer, M. Nieke, S. Schulze, C. Seidl, and I. Schaefer, "Multi-objective black-box test case selection for system testing," in *Proceedings of the Genetic and Evolutionary Computation Conference*, ser. GECCO '17. New York, NY, USA: Association for Computing Machinery, 2017, pp. 1311–1318.

[6] G. Lyon, *Nmap Network Scanning: Official Nmap Project Guide to Network Discovery and Security Scanning*. Insecure.Com, LLC, 2008.

[7] K.-T. Cho and K. G. Shin, "Fingerprinting electronic control units for vehicle intrusion detection," in *25th {USENIX} Security Symposium ({USENIX} Security 16)*, 2016, pp. 911–927.

[8] W. Choi, H. J. Jo, S. Woo, J. Y. Chun, J. Park, and D. H. Lee, "Identifying ecus using inimitable characteristics of signals in controller area networks," *IEEE Transactions on Vehicular Technology*, vol. 67, no. 6, pp. 4757–4770, 2018.

[9] M. Kneib and C. Huth, "On the fingerprinting of electronic control units using physical characteristics in controller area networks," *INFORMATIK 2017*, 2017.

[10] J. Huang, W. Albazrqaoe, and G. Xing, "Blueid: A practical system for bluetooth device identification," in *IEEE INFOCOM 2014 - IEEE Conference on Computer Communications*, 2014, pp. 2849–2857.

[11] O. Avatefipour, "Physical-fingerprinting of electronic control unit (ecu) based on machine learning algorithm for in-vehicle network communication protocol "can-bus"," Master's thesis, University of Michigan, Dearborn, Michigan, USA, 2017.

[12] D. R. Crow, S. R. Graham, and B. J. Borghetti, "Fingerprinting vehicles with can bus data samples," in *ICCWS 2020 15th International Conference on Cyber Warfare and Security*. Academic Conferences and publishing limited, 2020, p. 110.

[13] B. K. Aichernig, R. Bloem, M. Ebrahimi, M. Horn, F. Pernkopf, W. Roth, A. Rupp, M. Tappler, and M. Tranninger, "Learning a behavior model of hybrid systems through combining model-based testing and machine learning," in *Testing Software and Systems*, C. Gaston, N. Kosmatov, and P. Le Gall, Eds. Cham: Springer International Publishing, 2019, pp. 3–21.

[14] B. K. Aichernig, R. Bloem, M. Ebrahimi, M. Tappler, and J. Winter, "Automata learning for symbolic execution," in *2018 Formal Methods in Computer Aided Design (FMCAD)*, 2018, pp. 1–9.

[15] B. K. Aichernig, E. Muškardin, and A. Pferscher, "Learning-based fuzzing of iot message brokers," in *2021 IEEE 14th International Conference on Software Testing, Validation and Verification (ICST)*, 2021, yet to appear; preprint available at https://github.com/DES-Lab/Learning-Based-Fuzzing/blob/main/paper/aichernigMP-learning-based-fuzzing-of-iot-message-brokers-preprint.pdf.

[16] K. Hofer-Schmitz and B. Stojanović, "Towards formal verification of iot protocols: A review," *Computer Networks*, p. 107233, 2020.

[17] R. Ameur-Boulifa, F. Lugou, and L. Apvrille, "Sysml model transformation for safety and security analysis," in *Security and Safety Interplay of Intelligent Software Systems*, B. Hamid, B. Gallina, A. Shabtai, Y. Elovici, and J. Garcia-Alfaro, Eds. Cham: Springer International Publishing, 2019, pp. 35–49.

[18] B. Blanchet, B. Smyth, V. Cheval, and M. Sylvestre, "Proverif 2.00: automatic cryptographic protocol verifier, user manual and tutorial," Tech. Rep., 2018.

[19] M. Paulweber and K. Lebert, *Powertrain instrumentation and test systems*. Springer, 2016.

[20] S. Otten, T. Glock, C. P. Hohl, and E. Sax, "Model-based variant management in automotive systems engineering," in *2019 International Symposium on Systems Engineering (ISSE)*, 2019, pp. 1–7.

[21] D. E. Wilkins, *Practical planning: extending the classical AI planning paradigm*. Elsevier, 2014.

[22] A. Petrenko, F. Avellaneda, R. Groz, and C. Oriat, "Fsm inference and checking sequence construction are two sides of the same coin," *Software Quality Journal*, vol. 27, no. 2, pp. 651–674, 2019.

[23] D. Jackson, *Software Abstractions: logic, language, and analysis*. MIT press, 2012.

[24] ——, "Alloy: a language and tool for exploring software designs," *Communications of the ACM*, vol. 62, no. 9, pp. 66–76, 2019.

[25] C. Kruegel and Y. Shoshitaishvili, "Using static binary analysis to find vulnerabilities and backdoors in firmware," *Black Hat USA*, 2015.

[26] D. Song, D. Brumley, H. Yin, J. Caballero, I. Jager, M. G. Kang, Z. Liang, J. Newsome, P. Poosankam, and P. Saxena, "Bitblaze: A new approach to computer security via binary analysis," in *Information Systems Security*, R. Sekar and A. K. Pujari, Eds. Berlin, Heidelberg: Springer Berlin Heidelberg, 2008, pp. 1–25.

[27] N. Nethercote, "Dynamic binary analysis and instrumentation," University of Cambridge, Computer Laboratory, Tech. Rep., 2004.

[28] C. Phillips and L. P. Swiler, "A graph-based system for network-vulnerability analysis," in *Proceedings of the 1998 workshop on New security paradigms*. ACM, 1998, pp. 71–79.

[29] B. Schneier, "Attack trees," *Dr. Dobb's journal*, vol. 24, no. 12, pp. 21–29, 1999.

[30] C. A. Petri, "Kommunikation mit automaten," Ph.D. dissertation, Technische Universität Darmstadt, 1962.

[31] V. Varadharajan, "Petri net based modelling of information flow security requirements," in *[1990] Proceedings. The Computer Security Foundations Workshop III*, 1990, pp. 51–61.

[32] L. Yao, P. Dong, T. Zheng, H. Zhang, X. Du, and M. Guizani, "Network security analyzing and modeling based on Petri net and Attack tree for SDN," in *2016 International Conference on Computing, Networking and Communications (ICNC)*, 2016, pp. 1–5.

[33] S. Marksteiner and Z. Ma, "Approaching the automation of cyber security testing of connected vehicles," in *Proceedings of the Central European Cybersecurity Conference 2019*, ser. CECC 2019.    New York, NY, USA: ACM, 2019.